\documentclass[conference]{IEEEtran}

\ifCLASSINFOpdf
   \usepackage[pdftex]{graphicx}
\else
\fi
\usepackage{amsmath,amssymb,amsfonts}
\usepackage{braket}
\usepackage{xspace}
\usepackage[dvipsnames]{xcolor}

\begin{document}

\title{Hybrid Quantum Noise Approximation and Pattern Analysis on Parameterized  Component Distributions}

\author{\IEEEauthorblockN{Mouli Chakraborty\IEEEauthorrefmark{1}, Anshu Mukherjee\IEEEauthorrefmark{2}, Ioannis~Krikidis\IEEEauthorrefmark{3},
Avishek Nag\IEEEauthorrefmark{4},
Subhash Chandra\IEEEauthorrefmark{1}}\IEEEauthorblockA{\IEEEauthorrefmark{1}School of Natural Sciences,
Trinity College Dublin, The University of Dublin, College Green, Dublin 2, Ireland\\
\IEEEauthorrefmark{2}School of Electrical and Electronic Engineering,
University College Dublin, Belfield, Dublin 4, Ireland\\
\IEEEauthorrefmark{3}Department of Electrical and Computer Engineering, IRIDA Research Centre for Communication Technologies, \\University of Cyprus, Nicosia, Cyprus\\
\IEEEauthorrefmark{4}School of Computer Science,
University College Dublin, Belfield, Dublin 4, Ireland\\
Email: chakrabm@tcd.ie, anshu.mukherjee@ieee.org, krikidis.ioannis@ucy.ac.cy, avishek.nag@ucd.ie, schandra@tcd.ie}}

\maketitle

\begin{abstract}
Noise is a vital factor in determining the accuracy of processing the information of the quantum channel. One must consider classical noise effects associated with quantum noise sources for more realistic modelling of quantum channels. A hybrid quantum noise model incorporating both quantum Poisson noise and classical additive white Gaussian noise (AWGN) can be interpreted as an infinite mixture of Gaussians with weightage from the Poisson distribution. The entropy measure of this function is difficult to calculate. This research developed how the infinite mixture can be well approximated by a finite mixture distribution depending on the Poisson parametric setting compared to the number of mixture components. The mathematical analysis of the characterization of hybrid quantum noise has been demonstrated based on Gaussian and Poisson parametric analysis. This helps in the pattern analysis of the parametric values of the component distribution, and it also helps in the calculation of hybrid noise entropy to understand hybrid quantum noise better.
  \emph{}
\end{abstract}

\IEEEpeerreviewmaketitle


\begin{IEEEkeywords}
 Quantum noise, qubits, classical additive white Gaussian noise, Gaussian quantum channel, entropy. 
\end{IEEEkeywords}

\section{Introduction}

Quantum communication uses principles of quantum mechanics, such as superposition, entanglement, and interference, to transmit information securely. The fundamental element of such communication is the quantum bit, or qubit, which, unlike classical bits, is susceptible to various forms of noise such as decoherence \cite{seifi2024Clas_Noise_QuantumDecoherene}, dephasing, and photon loss due to its interaction with the environment and its intrinsic properties.  Entanglement-based quantum communication, such as quantum key distribution (QKD) for space communication, experiences a high loss during the terrestrial distribution of photons. This limits communications via the satellite links, where the loss is dominated by diffraction instead of absorption and scattering\cite{ecker2023entanglement-basedQKD}. These noise types can disrupt quantum states, leading to a degradation of information through the loss of key quantum properties like superposition and entanglement \cite{guo2023noiseEntanglement}. Accurately modelling these quantum channels requires detailed characterization of noise parameters to mitigate their effects and maintain the integrity of transmitted quantum information. Quantum noise, distinct from classical noise, significantly affects quantum channels and requires analysis through density matrices and quantum operations. This requires a hybrid noise model that integrates both quantum and classical noise, employing statistical methods such as the Poisson distribution to capture the realistic behaviour of quantum channels effectively \cite{fox2006quantum}. Quantum Poisson-distributed noises include disturbances like photon-shot noise, electron-shot noise, quantum projection noise, and spontaneous emission of photons from quantum states. These quantum noises can be combined with classical noises. This merging of noises highlights the interaction between quantum and classical noise sources in different scientific and engineering fields \cite{mouli2024}.

Modelling realistic quantum channels necessitates accounting for both quantum and classical noise sources due to their distinct characteristics and impacts on quantum communication systems. Quantum Poisson noise \cite{fox2006quantum} is crucially linked to the uncertainty principle and photon counting, where photons, often carriers of quantum states, arrive at random, creating discrete noise events. These events are fundamental to quantum physics and impact applications such as quantum information processing \cite{nielsen2010}. In contrast, classical additive white Gaussian noise (AWGN) \cite{haykin2001communication}, prevalent in many communication systems, features a continuous amplitude distribution, providing a contrasting noise profile with a zero mean and defined variance. Integrating quantum Poisson noise with classical AWGN captures random, discrete disruptions from quantum sources and continuous disturbances from classical sources. This hybrid noise model reflects the complex noise environment in quantum channels more accurately, facilitating the development of effective strategies to mitigate these interferences and enhance the reliability and security of quantum communications \cite{guo2023Gaussain_QuantumNoise}. 

In quantum channel modeling, accurately simulating quantum noise necessitates careful consideration of the Poisson and the Gaussian parameters. The Poisson parameter, \(\lambda\), is crucial as it quantifies the average photon count and captures the quantum mechanical nature of light, reflecting the random emission and detection processes in quantum optics \cite{li2023poisson,chen2024weighted}. On the other hand, Gaussian components offer a probabilistic approach to modeling noise in quantum channels \cite{reynolds2009gaussian}, effectively incorporating both quantum and classical noise effects. This integration allows for a comprehensive representation of various noise sources affecting quantum states during transmission. Together, these models improve understanding of quantum noise complexities, facilitating the development of advanced error correction techniques and communication protocols designed to counteract the diverse noise characteristics in quantum information transfer.

In \cite{mouli2024}, the hybrid quantum noise model originally proposed utilizes an approximation of an infinite mixture of Gaussian components, with mixing coefficients derived from a Poisson distribution. Meanwhile, in \cite{Mouli2024MLTQE}, machine learning-based models such as the Gaussian mixture model and expectation maximization have been proposed to optimize the achievable rate of a model of a hybrid quantum noise-aided system model. However, existing research omitted discussions on methods for approximating the primary probability density function (p.d.f.) with a secondary p.d.f. and also lacked formal proof establishing the new approximate function as a valid p.d.f. to work with. This research addresses these gaps by proving that the approximate function qualifies as a p.d.f. In addition, we introduce techniques for transitioning from an infinite mixture model to a finite number of mixture components, enhancing the practical applicability of the model.\\
The main contributions of this research work are listed as follows:
\begin{itemize}
	
	\item Our framework utilizes a hybrid quantum noise model that integrates quantum Poissonian noise with classical AWGN. This model is conceptualized as an infinite mixture of Gaussian distributions, with weights derived from Poisson distributions, accurately representing the noise encountered in quantum communication systems. A pivotal aspect of our study is the influence of the photon count, indicated by the Poisson parameter, on this noise model. We explore approximating this complex noise model with a finite mixture of Gaussian distributions weighted by the Poisson distribution.
	
	\item We employ a part of the work, such as in \cite{mouli2024}, which was on the hybrid quantum noise model based on an infinite Gaussian mixture weighted by Poisson distribution. However, our objective is to provide a method to simplify this model into a finite number of Gaussian mixtures, contingent on the Poisson parameter, to estimate the entropy of the hybrid quantum noise. This analysis has not been provided in the existing work \cite{mouli2024}. We investigate how the Poisson parameter interrelates with the number of Gaussian mixture components necessary to simulate the hybrid quantum noise's original probability density function. 
	
	\item Our results reveal a comparative assessment indicating that the Poisson parameter selection and the number of mixing coefficients are not arbitrary but interdependent. This research presents an analysis of this relationship and a comparative study that examines various values and the corresponding number of components of the Gaussian mixture.
\end{itemize}

Overall, our work presents the effect of Gaussian and Poisson parameters on hybrid quantum noise in emerging quantum communication. 

\textit{Notation:} In this research work, we use $tr(\cdot)$ and $\mathbf{T(\cdot)}$ to denote the trace of a matrix and {Trace-preserving} map, respectively. For a matrix $\mathbf{A}$, $\mathbf{A}^\dag$ and ${\mathbf{A}}^t$ represent the adjoint and transpose, respectively. We denote the complex conjugate of a vector $\boldsymbol{\nu}$ by using $\boldsymbol{\nu}^*$, and the tensor product is denoted by $\otimes$. We represent the Gaussian density by $\mathcal{N}$. For a quantum state $\boldsymbol{\psi}$, ket and bra are denoted by $\ket{\boldsymbol{\psi}}$ and $\bra{\boldsymbol{\psi}}$, respectively.


\section{System Model}

A qubit is expressed as a superposition of the basis states $\ket{\boldsymbol{0}}$ and $\ket{\boldsymbol{1}}$, mathematically represented as $\ket{\boldsymbol{\psi}} = \boldsymbol{\alpha}\ket{\boldsymbol{0}} + \boldsymbol{\beta}\ket{\boldsymbol{1}}$, where $\boldsymbol{\alpha}$ and $\boldsymbol{\beta}$ are complex numbers with the condition $|\boldsymbol{\alpha}|^2 + |\boldsymbol{\beta}|^2 = 1$. For a qubit in pure state $\ket{\boldsymbol{\psi}}$, the corresponding density matrix $\boldsymbol{\rho}$ is given by $\boldsymbol{\rho} = \ket{\boldsymbol{\psi}}\bra{\boldsymbol{\psi}} = \begin{bmatrix} |\boldsymbol{\alpha}|^2 & \boldsymbol{\alpha}^*\boldsymbol{\beta} \\ \boldsymbol{\alpha\beta}^* & |\boldsymbol{\beta}|^2 \end{bmatrix}$. Mixed states, on the other hand, are represented by a density matrix that is a probabilistic mixture of pure states $\ket{\boldsymbol{\psi}_i}$, mathematically expressed as $\boldsymbol{\rho} = \sum_i p_i \ket{\boldsymbol{\psi}_i}\bra{\boldsymbol{\psi}_i}$, ensuring $\sum_i p_i = 1$ for probability normalization.

In quantum mechanics, a quantum channel is an entirely positive trace-preserving map that transforms quantum states, represented as density matrices, within a Hilbert space. Transformation by any quantum channel can be modeled using an environment-assisted unitary operation, mathematically defined as $\boldsymbol{\rho} \mapsto \operatorname{tr}_E[\boldsymbol{U}(\boldsymbol{\rho} \otimes \boldsymbol{\rho}_E)\boldsymbol{U}^\dagger]$, where $\boldsymbol{U}$ is a unitary transformation and $\boldsymbol{\rho}_E$ is the state of the environment $E$.

The most general representation of a Gaussian quantum channel is described by the transformation of covariance matrices, $\boldsymbol{\gamma} \mapsto \boldsymbol{A}^t\boldsymbol{\gamma}\boldsymbol{A} + \boldsymbol{Z}$, where $\boldsymbol{A}$ can be any real matrix that affects transformations such as amplification, attenuation, and rotation in phase space, while $\boldsymbol{Z}$ represents a noise term, incorporating both quantum and classical components. This model demonstrates that Gaussian channels can add noise to a system so that the output remains a valid quantum state, which is essential for realistic quantum communication systems, where both quantum and classical noises are considered additive.

\subsection{Qubit Model}

In quantum mechanics, the geometric representation of a pure-state qubit can be depicted as a point on the Bloch sphere, identified by coordinates \((\theta, \phi)\) on a two-dimensional surface. This state can be mathematically expressed using a bivariate function \(\zeta(\theta, \phi)\). When considering the noise impacts, it is assumed that the qubit remains close to its original position on the sphere despite fluctuations, allowing for minor adjustments in \(\theta\) while \(\phi\) remains essentially constant. The manipulation involves slight rotations, modeled as \((\theta, \phi) \mapsto (\tilde{\theta}, \phi \pm \delta)\), with \(\tilde{\theta}\) spanning the full circle from 0 to \(2\pi\) and \(\delta\) being a small variation, indicating minimal change in the \(\phi\) coordinate.

Extending to a mixed-qubit state, the representation grows to three dimensions within the Bloch sphere, introducing a radial distance \(r\), leading to a vector point \((\theta, \phi, r)\). This is expressed through a multivariate function \(\tilde{\zeta}(\theta, \phi, r)\), where adjustments in noise conditions allow the qubit to shift slightly within defined limits, mathematically described as \((\theta, \phi, r) \mapsto (\tilde{\theta}, \phi \pm \delta_1, r \pm \delta_2)\). These small changes \(\delta_1\) and \(\delta_2\) are treated equally for simplification, and both \(\phi\) and \(r\) are considered stable compared to \(\theta\).

The theoretical treatment extends to imagining qubits navigating along a circular path on the sphere, which can be as simple as a single rotation or as complex as multiple rotations akin to a spiral. This conceptual extension allows the qubit to be approximated by a single scalar \(\theta\), simplifying its description to a one-dimensional random variable that effectively captures the probability distribution of quantum noise impacting the qubit. This approach reduces computational complexity while retaining the essence of the qubit's behavior under noise within the quantum framework.

\subsection{Quantum Noise Model}

In a practical scenario, quantum noise is indistinguishable from classical noise in quantum communication \cite{castelletto2003quantum}. It is difficult to detect pure quantum noise in a real-world system separately from classical noises; therefore, one should consider both quantum and classical noise when modeling a noisy quantum channel. Mathematically, the combination of classical AWGN and quantum Poisson noise is represented by a random variable  $Z$ with the relation $Z=Z_1 + Z_2$. Quantum noise $Z_1$ follows a Poisson distribution with parameter $\lambda$ \cite{fox2006quantum}, while classical AWGN $Z_2$ follows a Gaussian distribution with parameters $\mu_{Z_2}$ and $\sigma_{Z_2}$ \cite{haykin2001communication}. The hybrid quantum noise $Z$ is statistically described by considering it as the convolution product of  $Z_{1}$ and $Z_{2}$ as \cite{mouli2024},
\begin{equation}
\begin{split}f_{Z}(z) & =\sum_{i=0}^{\infty}w_{i}\mathcal{N}\big(z;\mu_{i}^{(z)},\,{\sigma_{i}^{(z)}}^{2}\big)\end{split}
,
\label{eq pdf of hybrid quantum noise1}
\end{equation}
where 
\begin{equation}
\mathcal{N}\big(z;\mu_{i}^{(z)},\,{\sigma_{i}^{(z)}}^{2}\big)=\frac{1}{\sigma_{Z_{2}}\sqrt{2\pi}}e^{-\frac{1}{2}\big(\frac{z-i-\mu_{Z_{2}}}{\sigma_{Z_{2}}}\big)^{2}},
\end{equation}
is a Gaussian density with mean $\mu_{i}^{(z)}=i+\mu_{Z_{2}}$, standard deviation (s.d.) $\sigma_{i}^{(z)}=\sigma_{Z_{2}}$, and the weightage $w_{i}=\frac{e^{-\lambda}\lambda^i}{i!}$ , $\sum_{i=0}^{\infty} w_{i}=1$, $w_{i}\geq 0 \quad  \forall i = 0,1,2 \cdots $.
Note that in the expression \eqref{eq pdf of hybrid quantum noise1}, the p.d.f. $f_Z(z)$ of hybrid quantum noise is an infinite mixture of Gaussians with weightage from the Poisson distribution. This infinite mixture is challenging to interpret in 1-dimension and multi-dimension as well. It is possible to express the p.d.f. $f_Z(z)$ of hybrid quantum noise in a finite sum of the distributions of the underlying components. This research aims to determine the number of components that, in combination, can approximate the actual distribution. 
Mathematically, the following function
\begin{equation}
  \begin{split}
     \tilde{f}_Z(z) & 
             = \sum_{i=0}^{R} 
             \frac{e^{-\lambda}\lambda^i}{i!}\frac{1}{\sigma_{Z_2}\sqrt{2\pi}}e^{-\frac{1}{2}\Big(\frac{z-i-\mu_{Z_2}}{\sigma_{Z_2}}\Big)^2}\\
           & = \sum_{i=0}^{R} w_{i} \mathcal{N}\Big(z;\mu_{i}^{(z)},    \,{\sigma_{i}^{(z)}}^{2}\Big) ,
    \label{eq approx pdf of hybrid quantum noise2} 
   \end{split}    
\end{equation}
will replace the p.d.f. in \eqref{eq pdf of hybrid quantum noise1} depending on certain aspects, where $R$ is the number of Gaussian components. Not every combination of values of $\lambda $ and $R$ can well approximate the p.d.f. in a finite domain of the function. We find the number of components for which the value of $R$ so that the function $\tilde{f}_Z(z)$ will serve the purpose of an approximation of $f_Z(z)$ in a finite domain.
In \eqref{eq approx pdf of hybrid quantum noise2}, the function $\tilde{f}_Z(z)$ is the Gaussian mixture in the scalar variable $z$, the coefficient $w_{i}\geq 0 \quad \forall i$ and $\sum_{i=0}^{R} w_{i} \approx 1$ for large $R$. We will study how this number of components will vary depending on the parametric value $\lambda$. The value of $\lambda$ is interrelated with the number of components $R$ in the finite summation expression \eqref{eq approx pdf of hybrid quantum noise2}, and this leads to the well-approx p.d.f of hybrid quantum noise. 
Once $\tilde{f}_Z(z)$ approximates $f_Z(z)$ in a finite domain, we will proceed to calculate the entropy of the hybrid quantum noise $Z$.
To evaluate the differential entropy of $Z$ one can consider the formula \cite{Huber2008}, \[ H(Z)=-\int_{\chi_{Z}}f_Z(z)\log_{2} f_{Z}(z)\,dz,\]
where $\chi_{Z}$ is the support of $f_{Z}$, i.e., the set on which $f_{Z}$ is nonzero.

\subsection{Hybrid Quantum Noise Model Analysis}

\begin{figure}[t]
\centering
\includegraphics[width=0.8\columnwidth]{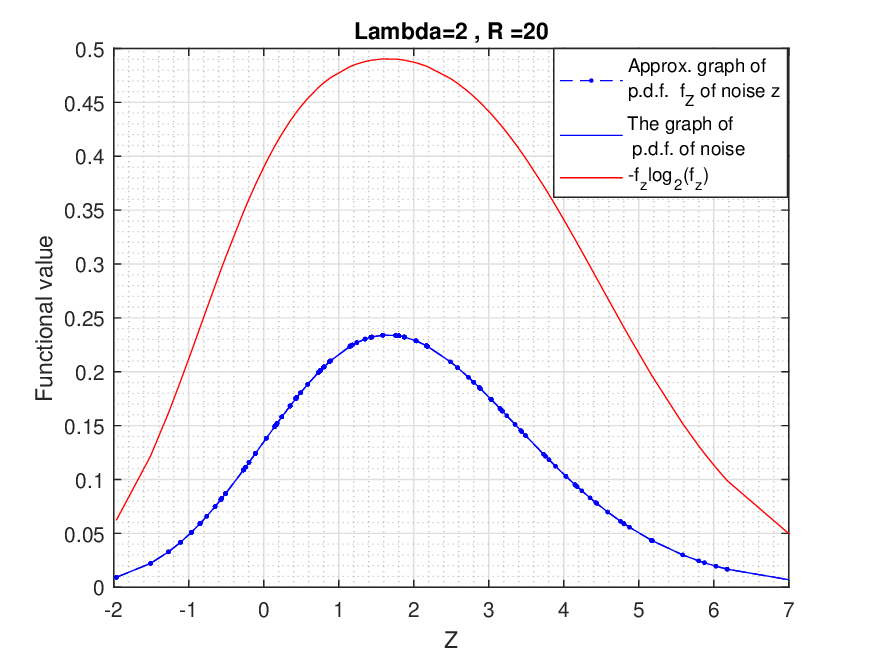}
\caption{Hybrid quantum noise model as a convolution of additive white Gaussian and quantum Poissonian noise with parameter $\lambda=2$ and number of mixing components $R=20$. } 
\vspace{-0.35cm}
\label{fig1}
\end{figure}

Consider the hybrid quantum noise function in \eqref{eq pdf of hybrid quantum noise1}, to replace the infinite mixture with a finite number of components, one should expect that the new function will approximate the original function in the domain of the function, mathematically stating that the function $\tilde{f}_Z$  can well approximate the original p.d.f $f_Z$  of hybrid quantum noise $Z$, depending on the parametric values of $\lambda$ and the number of mixing components $R$. To do so, we have to prove the following:
\begin{itemize}
    \item The function $\tilde{f}_Z$ approximately represents a p.d.f, that is, $\tilde{f}_Z(z) \geq 0 \quad \forall z \in \mathcal{D}_{Z} $ and
$\int_{\mathcal{D}_{Z}}\tilde{f}_Z(z)\,dz \approx 1 $ where $\mathcal{D}_{Z}$ is the domain of $f_{Z}$. 

\item $f_Z(z) $ and $ \tilde{f}_Z(z)$ are identical on the domain of the functions, that is  $f_Z(z) = \tilde{f}_Z(z)$ $\forall \ z\in \mathcal{D}_{Z} $ $\implies$ $\tilde{f}_Z$ will approximate well the original p.d.f of $f_Z$ within the domain of functions. Mathematically, the two functions $f: A \rightarrow B$ and $g: A \rightarrow B$ are identical if $f(a)=g(a)$ $\forall \ a\in A $.

\end{itemize}

It is important to note that, to do these, we need first to define the domain $\mathcal{D}_{Z}$  of the functions on which these functions can be replaced by each other. It is also essential to analyze how the parametric values of $\lambda$ and the number of mixing components $R$ are interdependent to approximate the p.d.f of noise $Z$—starting the analysis by generating the domain $\mathcal{D}_{Z}$ of the functions. In the simulation, MATLAB randomly selects sample spaces for Gaussian noise based on specific parameters (mean \& standard deviation) of the distribution, and depending on the parametric value of $\lambda$, the sample space for Poisson noise has been generated, returning a joint sample space for hybrid quantum noise $Z$. We define both functions $f_Z$ and $\tilde{f}_Z$ in the sample space mentioned so that both are in the same domain. This method is used due to the lack of experimental hybrid quantum noise data and a limitation on the availability of pre-generated quantum noise data. Hence, we have the domain for these functions in which the functions can be compared.

Secondly, to prove the function $\tilde{f}_Z$ as a p.d.f., let compute the following 
\begin{equation}
  \begin{split}
     \int_{\mathcal{D}_{Z}}\tilde{f}_Z(z)\,dz &  =\int_{\mathcal{D}_{Z}}\sum_{i=0}^{R} 
             \frac{e^{-\lambda}\lambda^i}{i!}\frac{1}{\sigma_{Z_2}\sqrt{2\pi}}e^{-\frac{1}{2}\Big(\frac{z-i-\mu_{Z_2}}{\sigma_{Z_2}}\Big)^2}dz \\
    & =  \sum_{i=0}^{R} \int_{\mathcal{D}_{Z}} \frac{e^{-\lambda}\lambda^i}{i!}\frac{1}{\sigma_{Z_2}\sqrt{2\pi}}e^{-\frac{1}{2}\Big(\frac{z-i-\mu_{Z_2}}{\sigma_{Z_2}}\Big)^2}dz \\
    & = \sum_{i=0}^{R} w_{i}\int_{\mathcal{D}_{Z}} \mathcal{N}\Big(z;\mu_{i}^{(z)}, \,{\sigma_{i}^{(z)}}^{2}\Big) dz \\
    & = \sum_{i=0}^{R} w_{i} = \sum_{i=0}^{R} \frac{e^{-\lambda}\lambda^i}{i!} \approx 1, 
    \label{eq the entropy of Gaussian mixtures}  
  \end{split}   
\end{equation}
as $\mathcal{N}\Big(z;\mu_{i}^{(z)}, \,{\sigma_{i}^{(z)}}^{2}\Big)$ is a Gaussian p.d.f. in random variable $Z$, hence $\int_{\mathcal{D}_{Z}}\mathcal{N}\Big(z;\mu_{i}^{(z)}, \,{\sigma_{i}^{(z)}}^{2}\Big)=1$, and the mixing coefficient $w_{i}$ coming from Poisson distribution, so $\sum_{i=0}^{R} w_{i} \approx 1$ depending on $\lambda$ and $R$. Also note that $\tilde{f}_Z(z) \geq 0 \quad \forall z \in \mathcal{D}_{Z} $ as $\frac{e^{-\lambda}\lambda^i}{i!} \geq 0$ being an exponential function and $\lambda \geq 0$ and also $\frac{1}{\sigma_{Z_2}\sqrt{2\pi}}e^{-\frac{1}{2}\Big(\frac{z-i-\mu_{Z_2}}{\sigma_{Z_2}}\Big)^2} \geq 0$ as $\frac{1}{\sigma_{Z_2}\sqrt{2\pi}} \geq 0$ and $e^{-\frac{1}{2}\Big(\frac{z-i-\mu_{Z_2}}{\sigma_{Z_2}}\Big)^2} \geq 0$. So, it is proved that $\tilde{f}_Z(z)$ satisfies the criteria of being a p.d.f. in the domain of the function $\mathcal{D}_{Z}$.

Third, by definition, $f_Z(z) = \tilde{f}_Z(z)$ $\forall \ z\in \mathcal{D}_{Z}$. Hence, it can be inferred that $\tilde{f}_Z$ can be an approximate p.d.f. depending on the parametric value of $\lambda$ and the mixing coefficient $R$. Thus, in the following paragraphs, we discuss how this approximation depends on $\lambda$ and the mixing coefficient $R$, involving interesting results in the following numerical section.

\section{Numerical Analysis}

\begin{figure}[t]
\centering
\includegraphics[width=0.8\columnwidth]{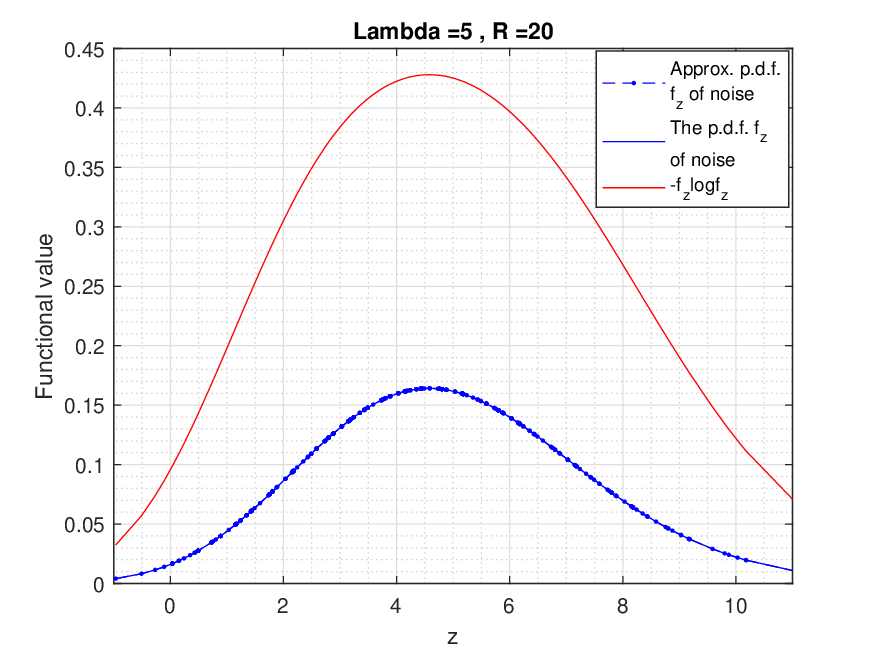}
\caption{Hybrid quantum noise model for $\lambda=5$ and number of mixing components $R=20$. Here the Poisson parameter \(\lambda\) is adequately balanced by a sufficient number of Gaussian components.} 
\vspace{-0.35cm}
\label{fig2}
\end{figure}

\begin{figure}[t]
\centering
\includegraphics[width=0.8\columnwidth]{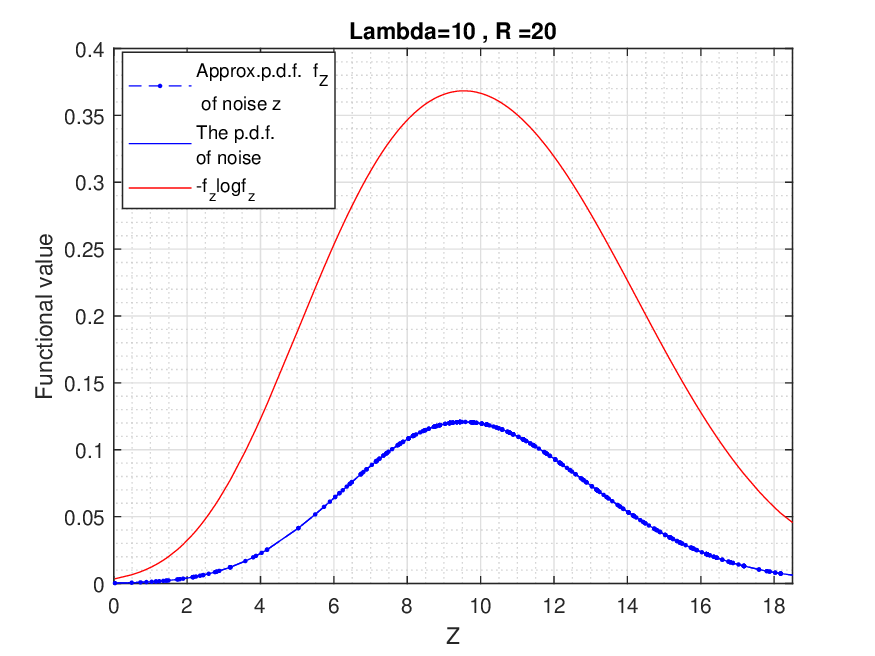}
\caption{Hybrid quantum noise model for  $\lambda=10$ and number of mixing components $R=20$. This approximation suffices to capture the actual behavior of the quantum noise and hence the the entropy of the hybrid quantum noise.} 
\vspace{-0.35cm}
\label{fig3}
\end{figure}

\begin{figure}[t]
\centering
\includegraphics[width=0.8\columnwidth]{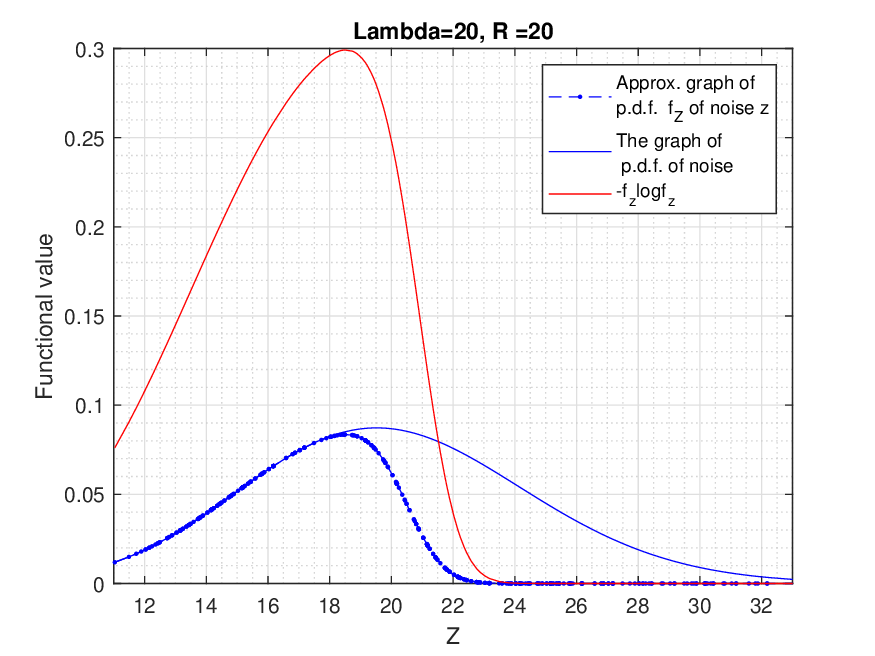}
\caption{Hybrid quantum noise model for $\lambda=20$ and number of mixing components $R=20$. Here \(\lambda\) is not adequately balanced by a sufficient number of Gaussian components.} 
\vspace{-0.35cm}
\label{fig4}
\end{figure}

\begin{figure}[t]
\centering
\includegraphics[width=0.8\columnwidth]{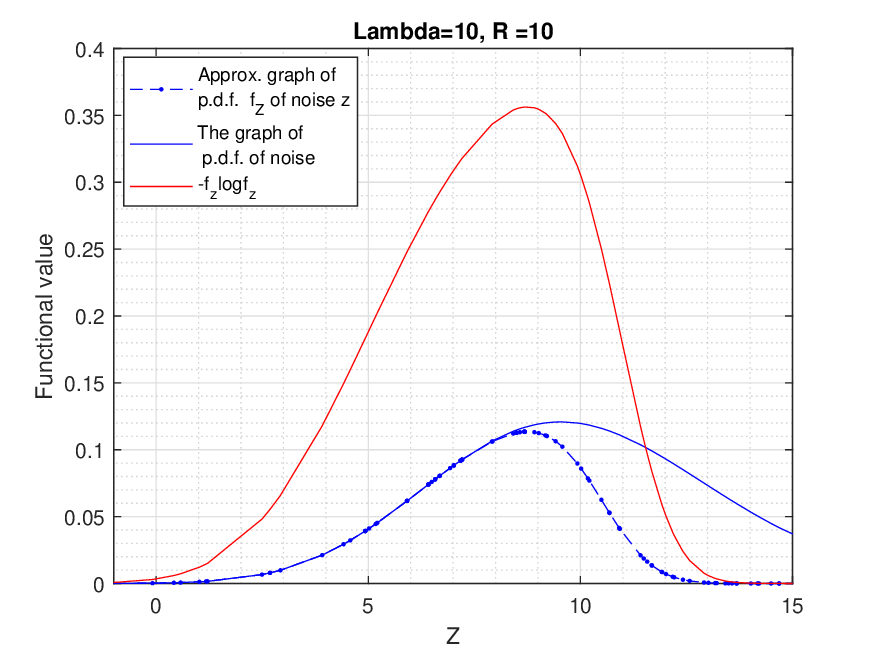}
\caption{Hybrid quantum noise model for $\lambda=10$ and number of mixing components $R=10$. This approximation fails to capture the actual behavior of the quantum noise.} 
\vspace{-0.35cm}
\label{fig5}
\end{figure}

\begin{figure}[t]
\centering
\includegraphics[width=0.8\columnwidth]{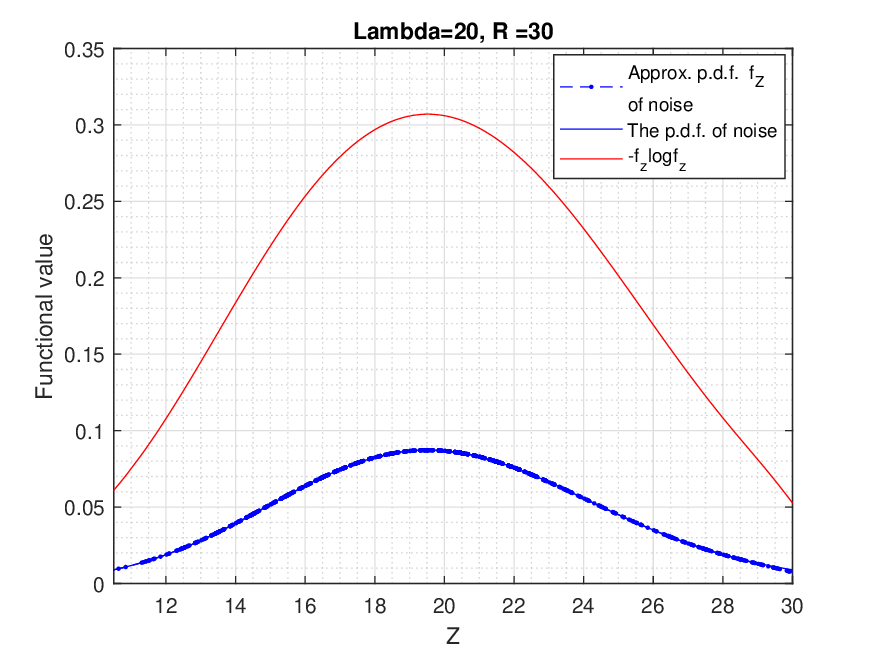}
\caption{Hybrid quantum noise model for $\lambda=20$ and number of mixing components $R=30$. Here, the Poisson parameter \(\lambda\) is adequately balanced by a sufficient number of Gaussian components; the approximation suffices to capture the actual behavior of the quantum noise and, hence the, the entropy of the hybrid quantum noise.} 
\vspace{-0.35cm}
\label{fig6}
\end{figure}

\subsection{Hybrid Quantum Noise Model Visualization}

The research focuses on modeling hybrid quantum noise using a finite number of Gaussian components, aiming to replicate the behavior of an infinite Gaussian mixture through discrete approximations. In MATLAB simulations, the p.d.f.s \(f_Z\) and \(\tilde{f}_Z\) of the hybrid quantum noise \(Z\) are compared, where they share the same domain and demonstrate equivalence across all parameters studied. These functions are analyzed under varying conditions of the Poisson parameter \(\lambda\) and the number of Gaussian components \(R\), with consistent Gaussian parameters \(\mu_{Z_2}=0\) and \(\sigma_{Z_2}=1\).

Through a series of figures, the relationship between \(\lambda\) and \(R\) is explored, providing information on these parameters' impact on the noise approximation's accuracy. In particular, Fig.~\ref{fig1} shows how well \(\tilde{f}_Z\) approximates \(f_Z\) for \(\lambda=2\) and \(R=20\), including an analysis of the entropy function that gives a detailed view of the noise characteristics in these settings. This methodological approach is consistent across various scenarios with different values of \(\lambda\), as seen in the subsequent figures Fig.~\ref{fig2} and Fig.~\ref{fig3}.

As \(\lambda\) increases, discrepancies in the approximation accuracy become apparent, mainly when \(\lambda\) equals \(R\), as shown in Fig.~\ref{fig4}  where the model underperforms. This observation suggests that the number of Gaussian components \(R\) must exceed \(\lambda\) to ensure a reliable approximation of the hybrid quantum noise p.d.f. This principle is further tested in Fig.~\ref{fig5} and Fig.~\ref{fig6}, where adjustments in \(R\) are made to improve approximation fidelity, demonstrating that increasing \(R\) relative to \(\lambda\) can significantly enhance the model's accuracy.

The study reveals that the parameter \(\lambda\), which represents the number of photons or the rate of quantum events per unit of time, crucially influences the fidelity of the noise model. If \(\lambda\) is not adequately balanced by a sufficient number of Gaussian components, the approximation fails to capture the actual behavior of the quantum noise. This finding underscores the necessity of carefully choosing \(R\) based on the values of \(\lambda\) to ensure that the Gaussian mixtures effectively approximate the complex dynamics of hybrid quantum noise.

In general, this research advances the understanding of quantum noise modeling by establishing critical parameters that govern the effective simulation of quantum channels in communication systems. It highlights the importance of parameter selection in achieving accurate noise models and lays a foundation for future explorations into more sophisticated quantum communication technologies. The conclusion drawn is that, for a precise simulation, the number of components of the Gaussian mixture must significantly exceed the Poisson parameter \(\lambda\), as succinctly expressed by the inequality \(\lambda < R\). This parameter interplay is vital for developing reliable quantum communication systems as it directly affects the integrity and performance of quantum information processing.

\section{Conclusion}

 This research work significantly contributes to this area by mathematically modeling hybrid quantum noise in practical communication channels. A novel aspect of this research is the calculation of the entropy of the hybrid noise, which provides deeper insight into the characteristics of the hybrid quantum noise. Its primary achievement is demonstrating how the p.d.f. of hybrid quantum noise can be effectively approximated using a finite number of component distributions. The study emphasizes the necessity of considering the Poisson parameter related to the number of mixing components in the hybrid quantum noise. It articulates the correlation between the control of flying photons (or the clicks on a Geiger counter, denoted by the parameter) and the number of mixing components in the hybrid quantum noise. This relationship is crucial to improving our understanding of practical noisy quantum channels, marking a significant advancement in the field.

\bibliographystyle{IEEEtran}
\bibliography{IEEEabrv,paper}
%



\end{document}